\documentstyle [preprint,revtex]{aps}

\begin{document}

\hfill {CEBAF-TH-93-04 / MIT-CTP-2187 / ORNL-CCIP-93-01}

\begin{title}
{
NN Core Interactions and Differential Cross Sections\\
from One Gluon Exchange\\
}
\end{title}

\author{T.Barnes}

\begin{instit}
Physics Division and Center for Computationally Intensive Physics\\
Oak Ridge National Laboratory, Oak Ridge, TN 37831-6373\\
and\\
Department of Physics\\
University of Tennessee, Knoxville, TN 37996-1200\\
\end{instit}

\author{S.Capstick}
\begin{instit}
Continuous Electron Beam Accelerator Facility\\
12000 Jefferson Ave., Newport News, VA 23606\\
\end{instit}

\author{M.D.Kovarik}

\begin{instit}
Physics Division and Center for Computationally Intensive Physics\\
Oak Ridge National Laboratory, Oak Ridge, TN 37831-6373\\
and\\
University of Tennessee Computer Center\\
University of Tennessee, Knoxville, TN 37996-0520\\
\end{instit}

\author{E.S.Swanson}

\begin{instit}
Center for Theoretical Physics\\
Laboratory of Nuclear Science and Department of Physics\\
Massachusetts Institute of Technology, Cambridge, MA 02139\\
\end{instit}

\begin{abstract}
We derive nonstrange baryon-baryon scattering amplitudes in the nonrelativistic
quark model using the ``quark Born diagram" formalism. This approach describes
the scattering as a single interaction, here the one-gluon-exchange (OGE)
spin-spin term followed
by constituent interchange, with external nonrelativistic baryon wavefunctions
attached to the scattering diagrams to incorporate higher-twist
wavefunction effects. The
short-range repulsive core in the NN interaction has previously been attributed
to this spin-spin interaction in the literature; we find that these
perturbative constituent-interchange diagrams do indeed predict repulsive
interactions in all I,S channels of the nucleon-nucleon system, and we compare
our results for the equivalent short-range potentials to the core potentials
found by other authors using nonperturbative methods. We also apply our
perturbative techniques to the N$\Delta$ and $\Delta\Delta$ systems: Some
$\Delta\Delta$ channels are found to have attractive core
potentials and may accommodate ``molecular" bound states near threshold.
Finally we use our Born formalism to calculate the NN differential
cross section, which we compare with experimental results for unpolarised
proton-proton elastic scattering. We find that several familiar features of the
experimental differential cross section are reproduced by our Born-order
result.

\end{abstract}

\centerline{February 1993}

\newpage

\section{Introduction}

The nucleon-nucleon interaction exhibits a strongly repulsive short-distance
core and a longer-ranged but much weaker attraction. Although there has been
evidence of the general features of this interaction for over fifty years
\cite{earlycore},
the physical mechanisms proposed as the origin of the interaction
have changed as our understanding of the strong interaction has progressed. In
1935 Yukawa \cite{Yukawa} suggested that the finite-ranged nuclear attraction
was due to the exchange of a massive, strongly-interacting meson, by analogy
with electron sharing as the origin of chemical forces. This hypothetical meson
was identified with the pion after its discovery, and as the lightest hadron it
certainly contributes the longest-ranged component of the nucleon-nucleon
strong force. The repulsive short-range core of the nucleon-nucleon interaction
was similarly attributed to the exchange of heavier mesons such as the $\omega$
after their discovery. These meson-exchange models have been elaborated
considerably since these original suggestions, and the most accurate
phenomenological descriptions of the nucleon-nucleon interaction at present are
meson exchange models \cite{OBE}, with parameters such as meson-nucleon
couplings fitted to experiment.

A literal attribution of the short-range repulsive core to
vector meson exchange, as opposed to a phenomenological parametrization,
of course involves a {\it non sequitur} \cite{Isgur,siemens}: Since the
nucleons
have radii of $\approx 0.8$ fm, and the range of the vector-exchange force is
$1/m_{\omega}\approx 0.2$ fm, one would have to superimpose the nucleon
wavefunctions to reach the appropriate internucleon separations. The picture of
distinct nucleons exchanging a physical $\omega$ meson at such a small
separation is clearly a fiction, and a realistic description of the short-range
core interaction requires a treatment of the quark wavefunctions of
the interacting nucleons and a Hamiltonian which involves quark and gluon
degrees of freedom.

Since the development of QCD as the theory of the strong interaction there have
been many studies of the NN interaction in terms of quarks and gluons.
Most have employed the nonrelativistic quark potential
model, although some early work used the MIT bag model \cite{detar}.
Many studies of this and other multiquark systems were
fundamentally flawed
due to an inadequate treatment of the color degree of freedom, or due to
assumptions that imposed confinement on the entire multiquark system; a
discussion of these problems is given by Isgur \cite{Isgur}.
It now appears that a pairwise $\lambda \cdot \lambda$
color interaction
\cite{Nambu,dRGG,Schnitzer} together with a sufficiently
general spatial wavefunction that allows dissociation into color singlets
provides a sufficiently
realistic description of color forces in a multiquark system.
The NN references summarised here all assume the
$\lambda \cdot \lambda$ form, and it
has also become standard to employ a quadratic or linear confinement
potential. Finally, the spin-spin hyperfine term is incorporated in
all these references,
as it apparently makes a dominant contribution to the NN core interaction.
Several groups have included other terms from the Breit-Fermi interaction
such as the hyperfine-tensor and spin-orbit interactions.

Attempts to describe the NN interaction using quark potential models date from
the work of Liberman in 1977 \cite{Liberman}, who calculated the adiabatic
response of the six quark system to variations in the interbaryon coordinate.
The resulting effective potential had a repulsive core with a weak
intermediate-range attraction. Liberman concluded that the repulsive core was
predominantly due to a combination of the Pauli principle and the contact
hyperfine term. The same conclusion was reached by Neudatchin and collaborators
\cite{Russ}, also within the adiabatic approximation. Harvey \cite{Harvey}
continued this adiabatic approach in a generalized calculation of the effective
NN potential; he noted that SU(4) spin-isospin symmetry required that NN,
$\Delta \Delta$, and ``hidden color '' $(qqq)_8 \otimes (qqq)_8$ states be
included in the calculation. On incorporating these states, he found that the
repulsive core was strongly suppressed. It is now
widely believed that the absence of
the repulsive core in improved calculations is an artifact of the adiabatic
approximation \cite{FFLS,HLL}.

The usual method for improving on
the adiabatic approximation in the $q^6$ system is to employ the
resonating-group method. This involves expanding the wavefunction of
the system in a basis which describes system subclusters multiplied by
unknown functions of the intercluster coordinates. One then solves the
resulting coupled integro-differential equations using various numerical
techniques.
Baryon subcluster wavefunctions are usually
taken to be simple Gaussians, and the coupled system is truncated
at the NN, NN + $\Delta \Delta$, or NN + $\Delta \Delta$ + hidden color levels.
The first applications of the resonating-group method
to the NN system were given by Warke and Shanker \cite{WS},
Oka and Yazaki \cite{OY} and Ribeiro \cite{rib}. All groups found a
repulsive core, which was dominantly due to the Pauli principle and the color
hyperfine term.
The $\Delta \Delta$ and
hidden color channels were found to make
only small contributions to the hard-core
S-wave phase shifts.
These studies found that several contributions including the
Coulomb and confinement kernels approximately
cancelled, leaving the spin-spin OGE hyperfine term as the dominant
interaction.
The resonating-group approach has been extended to include
more channels, strange quarks, effective one-boson-exchange long-range
potentials, and virtual excitations of the quark wavefunctions.
In particular, Koike
\cite{Koike} has applied these techniques to the ``flip-flop" model
\cite{flop} (which eliminates long-range color van der Waals forces),
supplemented by an effective meson-exchange potential, and
Cao and Kisslinger \cite{CK} have developed a relativised
resonating-group formalism and applied it to the
determination of equivalent potentials
and low-L phase shifts in a model which incorporates OGE and meson exchange
forces.
Both these references find reasonably good
agreement with experimental low-energy NN phase shifts.
A summary of work in this field to 1989 has been given by Shimizu \cite{Shim}.

Maltman and Isgur \cite{MI} have performed a detailed variational calculation
of the ground state properties of the deuteron using a $\lambda \cdot \lambda$
quark-quark interaction with the full OGE color Breit-Fermi interaction, a
quadratic confinement term, and a phenomenological one-pion-exchange potential.
In contrast to typical resonating-group calculations, they allowed spatial
excitations within the clusters. They found a repulsive core, and noted that
the admixture of P-wave color octet clusters significantly increased the range
and depth of the intermediate-range attraction. Their results for the deuteron
binding energy, RMS radius, quadrupole moment, and magnetic moment all agreed
well with experiment.

Although the origin of the nucleon-nucleon force at the QCD level is now
reasonably well understood, the resonating-group and variational techniques
which have been employed in this work are rather intricate and
require considerable theoretical
effort, and usually lead to numerical rather than analytical results. As these
techniques are best suited to the determination of ground state properties,
topics such as resonance production and scattering cross sections
at higher energies have received little attention in these quark model studies.

In recent work we have investigated the possibility that these low-energy
nonresonant hadronic scattering amplitudes may actually be dominated by simple
perturbative processes; if so, it may be possible to derive useful estimates of
these amplitudes using a much simpler approach. A complementary possibility of
perturbative dominance of hadronic scattering processes at high energies
through constituent-interchange mechanisms has been investigated by
theorists almost since the development of QCD \cite{glennys}. Results for
elastic hadron-hadron scattering amplitudes, in particular
the asymptotic $Q^2$ dependence of fixed-angle scattering,
have recently been presented by Botts and Sterman \cite{Botts}.

Concerns regarding the range of validity of the high-energy
perturbative QCD studies have
been expressed by Isgur and Llewellyn-Smith \cite{ILS}, who suggest that
higher-twist hadron wavefunction effects may actually dominate perturbative
QCD contributions at experimentally accessible energies. We have explicitly
incorporated these wavefunction effects in our study of a
constituent-interchange scattering mechanism in the nonrelativistic quark
potential model. We calculate the hadron-hadron scattering amplitudes which
follow from one gluon exchange followed by constituent interchange (quark line
interchange is required at lowest order in $\alpha_s$ to restore color singlet
final states), with nonrelativistic quark model wavefunctions attached to the
external lines. This OGE+CI mechanism may be dominant in processes in which
$q\bar q$ annihilation is forbidden for the valence wavefunctions. We have
applied this description of scattering to elastic
I=2 $\pi\pi$ \cite{BS} and I=3/2
K$\pi$ \cite{BSW} reactions and found excellent agreement with the
experimental S-wave phase shifts given standard quark model parameters.
Related approaches to calculating meson-meson scattering amplitudes which
iterate this quark-gluon mechanism
have been discussed in the
literature \cite{othermes}.
These Born-order
techniques have also been applied to vector-vector meson systems
\cite{Swanson}, and lead to interesting predictions of vector-vector molecule
bound states in certain channels \cite{DSB}. In the vector-vector system the
hyperfine interaction apparently does not dominate the scattering amplitude,
unlike the pseudoscalar-pseudoscalar and NN systems.
More recently we applied the
quark Born formalism to KN scattering \cite{BSKN}, which is also free of $q\bar
q$ annihilation at the valence quark level. We found satisfactory agreement
with the experimental S-wave KN scattering lengths, although I=0 is not yet
very well determined experimentally. The KN S-wave phase shifts at higher
energies, however, are not well described; they require stronger high-momentum
components in the nucleon wavefunction than are present in the single Gaussian
forms we assumed. The higher-L KN partial waves, especially the P-waves, show
evidence of a spin-orbit interaction which does not arise in single-channel
spin-spin scattering, which has not yet been adequately explained in the
literature.

The next level of complexity in Hilbert space is the $q^6$ baryon-baryon
sector. Since this system is free of annihilation at the valence level, and the
spin-spin hyperfine term has already been established as the dominant
interaction underlying the core repulsion, derivation of the NN core
interaction is an important test of the quark Born formalism. Here we derive
the nucleon-nucleon interaction predicted by the OGE spin-spin term using quark
Born diagrams, and show that the predicted core interaction is indeed strongly
repulsive in all four spin and isospin channels. Low energy Born-equivalent NN
core potentials are also derived and compared to previous results. We then
consider other nonstrange baryons and derive the N$\Delta$ and $\Delta\Delta$
short-range interactions; some of these are found to be attractive, and we
investigate the possibility that these channels might support dibaryon molecule
bound states. Some of our results for attractive $\Delta\Delta$ channels are
consistent with the previous conclusions of Maltman \cite{Maltman}. Finally we
derive the elastic NN differential cross section predicted by our quark Born
formalism and find that some familiar experimental features of the high-energy
elastic proton-proton differential cross section are evident in our results.

\section{derivation of scattering amplitudes}

\noindent
{\it a) Hamiltonian and hadron states}
\vskip 0.2cm

In the quark Born diagram formalism we derive the matrix element of the
interaction Hamiltonian between quarks in incoming hadron states to leading
Born order, which is then used to calculate scattering observables. We factor
out the overall momentum conserving delta function and then derive the
remaining matrix element, which we call
$h_{fi}$;
\begin{equation}
{}_f\langle BB'  | H_{scat} | BB'  \rangle_i \equiv h_{fi} \ \delta(\vec P_f -
\vec P_i) \ .
\label{eq:hfi}
\end{equation}
Since the hadron state normalizations we will introduce are identical to those
used in our previous study of I=2 $\pi\pi$ scattering \cite{BS} we can use the
relations between the scattering matrix element $h_{fi}$ and the phase shifts
and cross sections given there. The details of our diagrammatic
procedure for determining $h_{fi}$ are described elsewhere \cite{BS,BSW,BSKN};
here we shall simply recall some basic points and then give our results.

For baryon-baryon
scattering we shall follow previous studies
\cite{Isgur,Liberman,Russ,Harvey,FFLS,HLL,WS,OY,rib,Koike,flop,Shim,MI}
and
assume that the dominant part of the core interaction derives from the
spin-spin color
hyperfine term,
\begin{equation}
H_{scat} = \sum_{a,i<j} \;
\bigg{[}
-{8 \pi \alpha_s\over 3 m_i m_j}\; \delta(\vec r_{ij})
\bigg{]}
\;
\bigg{[}
\vec S_i \cdot \vec S_j
\bigg{]}
\;
\bigg{[}
{\cal F}^{\, a}_i \cdot {\cal F}^{\, a}_j
\bigg{]} \ ,
\label{eq:HI}
\end{equation}
\noindent
where ${\cal F}^{\, a}_i$ is the color matrix $\lambda^a/2$ for quark $i$.
The baryon color wavefunctions are the usual color singlets,
\begin{equation}
|baryon\rangle = \sum_{i,j,k=1,3}{1\over \sqrt{6}} \;
\epsilon_{ijk} \
|ijk\rangle  \ .
\end{equation}
Our spin-flavor states for the meson and baryon are the usual SU(6) states, but
as explained in reference \cite{BSKN} we find it convenient to write these
states using field theory conventions rather than in the usual quark model
form. The quark model conventions show explicit exchange symmetry by assigning
a fixed location in the state vector to each quark. The field theoretic
convention greatly reduces the number of terms encountered in our scattering
matrix elements; for example, the proton state in field theory conventions has
only 2 terms instead of the usual 9 for quark model states, so PP$\to$PP
elastic scattering involves only 16 terms, far fewer than the 6561 we would
encounter with the usual quark model conventions.
As examples, the
orthonormal $S_z=3/2$ $|\Delta^+_{3/2}\rangle$ and $S_z=1/2$ $|P_{1/2}\rangle$
states in field theory conventions are
\begin{equation}
| \Delta^+_{3/2} \rangle = {1\over \sqrt{2}} |u_+ u_+ d_+ \rangle
\end{equation}
and
\begin{equation}
| P_{1/2} \rangle =
\sqrt{2\over 3}\ \bigg\{ {|u_+ u_+ d_- \rangle \over \sqrt{2}} \bigg\}
-
\sqrt{1\over 3}\ |u_+ u_- d_+ \rangle  \ .
\end{equation}
The other baryon states considered in this paper can be derived from these
by application of spin and isospin raising and lowering operators.

We shall quote general results for baryon-baryon scattering amplitudes with
arbitrary spatial wavefunctions attached to the external lines
and then specialize to single-Gaussian forms to derive representative
closed-form results. The general spatial baryon wavefunction we assume is of
the form
\begin{equation}
\Phi_{baryon}(\vec p_1,\vec p_2,\vec p_3;\vec P_{tot}) =
\phi_{baryon}(\vec p_1,\vec p_2,\vec p_3) \;
\delta(\vec P_{tot} - \vec p_1 - \vec p_2 - \vec p_3 )
\end{equation}
with a normalization given by
\begin{displaymath}
\langle \Phi_{baryon}(\vec P'_{tot}) | \Phi_{baryon}(\vec P_{tot}) \rangle
\phantom{yyyyyyyyyyyyyyyyyyyyyyyyyyyyyyy}
\end{displaymath}
\begin{displaymath}
= \int \!\! \int \!\! \int \!\!  \int \!\!  \int \!\!  \int \,
d\vec p_1 \, d\vec p_2 \, d\vec p_3 \,
d\vec p_1\,' \, \vec p_2\,' \, d\vec p_3\,' \,
\Phi^*_{baryon}(\vec p_1\,',\vec p_2\,',\vec p_3\,';\vec P'_{tot})
\Phi_{baryon}(\vec p_1,\vec p_2,\vec p_3;\vec P_{tot})
\end{displaymath}
\begin{equation}
= \delta(\vec P_{tot} - \vec P'_{tot}) \ .
\end{equation}

The standard quark model single-Gaussian baryon wavefunction we shall use for
illustration is
\begin{equation}
\phi_{baryon}(\vec p_1,\vec p_2,\vec p_3) = {3^{3/4} \over \pi^{3/2}
\alpha^3 }
\exp
\bigg\{
-
{
(
 \vec p_1^{\, 2}
+\vec p_2^{\, 2}
+\vec p_3^{\, 2}
- \vec p_1 \cdot \vec p_2
- \vec p_2 \cdot \vec p_3
- \vec p_3 \cdot \vec p_1
)
\over
3 \alpha^2
}
\bigg\} \ .
\label{eq:wfn}
\end{equation}
Oscillator parameter values of $0.25 \; {\rm GeV} \leq \alpha \leq 0.42 \; {\rm
GeV}$ have been used in the quark model literature on baryon spectroscopy, as
we will discuss subsequently.

\vskip 0.2cm
\noindent
{\it b) Baryon-baryon scattering amplitudes}
\vskip 0.2cm

By analogy with our study of KN scattering \cite{BSKN} (see especially section
IIb) we first write a generic scattering diagram with initial and final
baryon-baryon
states. We then connect the initial and final quark lines in all ways
consistent with flavor conservation; for example
$\Delta^{++}\Delta^{++}$ elastic
scattering has 6!=720 quark line diagrams. These may be
grouped into four sets in which the number of $q\bar q$ pairs which cross in
t-channel is zero, one, two or three. We then generate scattering diagrams by
inserting one-gluon-exchange interactions between all pairs of initial quarks
in different initial baryons; this gives nine times as many scattering diagrams
as we had quark line diagrams. Many of these diagrams are trivially zero; these
include the zero-pair-interchange and three-pair-interchange diagrams, which
vanish due to color. The nonzero scattering diagrams may be related to a small
``reduced set" of diagrams by permutation of external lines, which leaves a
diagram invariant. In baryon-baryon scattering this reduced set contains
eight independent diagrams, which are shown below.
\vskip 1cm
%*******************************************************************
%d 1
\setlength{\unitlength}{1.6pt}

\begin{picture}(320,70)(30,-5)
\put(70,28) {\makebox(0,0)[1]{ $D_1$ } }
\put(100,28) {\makebox(0,0)[1]{ = } }
\put(280,28) {\makebox(0,0)[1]{(9)} }
\put(230,-5) {\makebox(0,0)[1]{,} }

\put(116,55) {\makebox(0,0)[1]{ $A\Bigg\{ $    } }
\put(116,05) {\makebox(0,0)[1]{ $B\Bigg\{ $    } }
\put(224,55) {\makebox(0,0)[1]{ $\Bigg\} C$    } }
\put(224,05) {\makebox(0,0)[1]{ $\Bigg\} D$    } }

\put(120,0){

\begin{picture}(75,60)(0,0)

\multiput(10,65)(70,0){2}{\vector(1,0){5}}
\multiput(10,55)(70,0){2}{\vector(1,0){5}}
\multiput(10,45)(70,0){2}{\vector(1,0){5}}
\multiput(10,15)(70,0){2}{\vector(1,0){5}}
\multiput(10, 5)(70,0){2}{\vector(1,0){5}}
\multiput(10,-5)(70,0){2}{\vector(1,0){5}}
%
% full lines
\put(5,65){\line(1,0){85}}
\put(5,55){\line(1,0){85}}
\put(5,45){\line(1,0){35}}
\put(5,15){\line(1,0){35}}
\put(5,5){\line(1,0){85}}
\put(5,-5){\line(1,0){85}}

\put(60,45){\line(1,0){30}}
\put(60,15){\line(1,0){30}}

\put(40,15){\line(2,3){20}}
\put(40,45){\line(2,-3){20}}

% OGE between (25,15) and (25,45) (q and q1)
\put(25,15){\dashbox{2}(0,30){}}
\multiput(25,15)(0,30){2}{\circle*{2}}

\end{picture}
}

\end{picture}

%*******************************************************************
\vspace{1.0cm}
%*******************************************************************
%d 2
\setlength{\unitlength}{1.6pt}

\begin{picture}(320,70)(30,-5)
\put(70,28) {\makebox(0,0)[1]{ $D_2$ } }
\put(100,28) {\makebox(0,0)[1]{ = } }
\put(280,28) {\makebox(0,0)[1]{(10)} }
\put(220,-5) {\makebox(0,0)[1]{,} }

\put(120,0){

\begin{picture}(75,60)(0,0)

\multiput(10,65)(70,0){2}{\vector(1,0){5}}
\multiput(10,55)(70,0){2}{\vector(1,0){5}}
\multiput(10,45)(70,0){2}{\vector(1,0){5}}
\multiput(10,15)(70,0){2}{\vector(1,0){5}}
\multiput(10, 5)(70,0){2}{\vector(1,0){5}}
\multiput(10,-5)(70,0){2}{\vector(1,0){5}}
%
% full lines
\put(5,65){\line(1,0){85}}
\put(5,55){\line(1,0){85}}
\put(5,45){\line(1,0){35}}
\put(5,15){\line(1,0){35}}
\put(5,5){\line(1,0){85}}
\put(5,-5){\line(1,0){85}}

\put(60,45){\line(1,0){30}}
\put(60,15){\line(1,0){30}}

\put(40,15){\line(2,3){20}}
\put(40,45){\line(2,-3){20}}

% OGE between (25,15) and (25,45) (q and q1)
\put(25,15){\dashbox{2}(0,40){}}
\multiput(25,15)(0,40){2}{\circle*{2}}

\end{picture}
}

\end{picture}

%*******************************************************************
\vspace{1.0cm}
%*******************************************************************
%d 3
\setlength{\unitlength}{1.6pt}

\begin{picture}(320,70)(30,-5)
\put(70,28) {\makebox(0,0)[1]{ $D_3$ } }
\put(100,28) {\makebox(0,0)[1]{ = } }
\put(280,28) {\makebox(0,0)[1]{(11)} }
\put(220,-5) {\makebox(0,0)[1]{,} }

\put(120,0){

\begin{picture}(75,60)(0,0)

\multiput(10,65)(70,0){2}{\vector(1,0){5}}
\multiput(10,55)(70,0){2}{\vector(1,0){5}}
\multiput(10,45)(70,0){2}{\vector(1,0){5}}
\multiput(10,15)(70,0){2}{\vector(1,0){5}}
\multiput(10, 5)(70,0){2}{\vector(1,0){5}}
\multiput(10,-5)(70,0){2}{\vector(1,0){5}}
%
% full lines
\put(5,65){\line(1,0){85}}
\put(5,55){\line(1,0){85}}
\put(5,45){\line(1,0){35}}
\put(5,15){\line(1,0){35}}
\put(5,5){\line(1,0){85}}
\put(5,-5){\line(1,0){85}}

\put(60,45){\line(1,0){30}}
\put(60,15){\line(1,0){30}}

\put(40,15){\line(2,3){20}}
\put(40,45){\line(2,-3){20}}

% OGE between (25,15) and (25,45) (q and q1)
\put(25, 5){\dashbox{2}(0,40){}}
\multiput(25, 5)(0,40){2}{\circle*{2}}

\end{picture}
}

\end{picture}

%*******************************************************************
\vspace{1.0cm}
%*******************************************************************
%d 4
\setlength{\unitlength}{1.6pt}

\begin{picture}(320,70)(30,-5)
\put(70,28) {\makebox(0,0)[1]{ $D_4$ } }
\put(100,28) {\makebox(0,0)[1]{ = } }
\put(280,28) {\makebox(0,0)[1]{(12)} }
\put(220,-5) {\makebox(0,0)[1]{,} }

\put(120,0){

\begin{picture}(75,60)(0,0)

\multiput(10,65)(70,0){2}{\vector(1,0){5}}
\multiput(10,55)(70,0){2}{\vector(1,0){5}}
\multiput(10,45)(70,0){2}{\vector(1,0){5}}
\multiput(10,15)(70,0){2}{\vector(1,0){5}}
\multiput(10, 5)(70,0){2}{\vector(1,0){5}}
\multiput(10,-5)(70,0){2}{\vector(1,0){5}}
%
% full lines
\put(5,65){\line(1,0){85}}
\put(5,55){\line(1,0){85}}
\put(5,45){\line(1,0){35}}
\put(5,15){\line(1,0){35}}
\put(5,5){\line(1,0){85}}
\put(5,-5){\line(1,0){85}}

\put(60,45){\line(1,0){30}}
\put(60,15){\line(1,0){30}}

\put(40,15){\line(2,3){20}}
\put(40,45){\line(2,-3){20}}

% OGE between (25,15) and (25,45) (q and q1)
\put(25, 5){\dashbox{2}(0,50){}}
\multiput(25, 5)(0,50){2}{\circle*{2}}

\end{picture}
}

\end{picture}

%*******************************************************************
\vspace{1.0cm}

%*******************************************************************
%d 5

\setlength{\unitlength}{1.6pt}

\begin{picture}(320,70)(30,-5)
\put(70,28) {\makebox(0,0)[1]{ $D_5$ } }
\put(100,28) {\makebox(0,0)[1]{ = } }
\put(280,28) {\makebox(0,0)[1]{(13)} }
\put(220,-5) {\makebox(0,0)[1]{,} }

\put(120,0){

\begin{picture}(75,60)(0,0)

\multiput(10,65)(70,0){2}{\vector(1,0){5}}
\multiput(10,55)(70,0){2}{\vector(1,0){5}}
\multiput(10,45)(70,0){2}{\vector(1,0){5}}
\multiput(10,15)(70,0){2}{\vector(1,0){5}}
\multiput(10, 5)(70,0){2}{\vector(1,0){5}}
\multiput(10,-5)(70,0){2}{\vector(1,0){5}}
%
% full lines
\put(5,65){\line(1,0){85}}
\put(5,55){\line(1,0){35}}
\put(5,45){\line(1,0){25}}
\put(5,15){\line(1,0){25}}
\put(5,5){\line(1,0){35}}
\put(5,-5){\line(1,0){85}}

\put(60,55){\line(1,0){30}}
\put(60, 5){\line(1,0){30}}

\put(70,45){\line(1,0){20}}
\put(70,15){\line(1,0){20}}

\put(30,15){\line(3,4){30}}
\put(30,45){\line(3,-4){30}}

\put(40, 5){\line(3,4){30}}
\put(40,55){\line(3,-4){30}}

% OGE between (25,15) and (25,45) (q and q1)
\put(25,-5){\dashbox{2}(0,70){}}
\multiput(25,-5)(0,70){2}{\circle*{2}}

\end{picture}
}

\end{picture}

%*******************************************************************
\vspace{1.0cm}

%*******************************************************************
%d 6

\setlength{\unitlength}{1.6pt}

\begin{picture}(320,70)(30,-5)
\put(70,28) {\makebox(0,0)[1]{ $D_6$ } }
\put(100,28) {\makebox(0,0)[1]{ = } }
\put(280,28) {\makebox(0,0)[1]{(14)} }
\put(220,-5) {\makebox(0,0)[1]{,} }

\put(120,0){

\begin{picture}(75,60)(0,0)

\multiput(10,65)(70,0){2}{\vector(1,0){5}}
\multiput(10,55)(70,0){2}{\vector(1,0){5}}
\multiput(10,45)(70,0){2}{\vector(1,0){5}}
\multiput(10,15)(70,0){2}{\vector(1,0){5}}
\multiput(10, 5)(70,0){2}{\vector(1,0){5}}
\multiput(10,-5)(70,0){2}{\vector(1,0){5}}
%
% full lines
\put(5,65){\line(1,0){85}}
\put(5,55){\line(1,0){35}}
\put(5,45){\line(1,0){25}}
\put(5,15){\line(1,0){25}}
\put(5,5){\line(1,0){35}}
\put(5,-5){\line(1,0){85}}

\put(60,55){\line(1,0){30}}
\put(60, 5){\line(1,0){30}}

\put(70,45){\line(1,0){20}}
\put(70,15){\line(1,0){20}}

\put(30,15){\line(3,4){30}}
\put(30,45){\line(3,-4){30}}

\put(40, 5){\line(3,4){30}}
\put(40,55){\line(3,-4){30}}

% OGE between (25,15) and (25,45) (q and q1)
\put(25,-5){\dashbox{2}(0,50){}}
\multiput(25,-5)(0,50){2}{\circle*{2}}

\end{picture}
}

\end{picture}

%*******************************************************************
\vspace{1.0cm}
%*******************************************************************
%d 7

\setlength{\unitlength}{1.6pt}

\begin{picture}(320,70)(30,-5)
\put(70,28) {\makebox(0,0)[1]{ $D_7$ } }
\put(100,28) {\makebox(0,0)[1]{ = } }
\put(280,28) {\makebox(0,0)[1]{(15)} }
\put(220,-5) {\makebox(0,0)[1]{,} }

\put(120,0){

\begin{picture}(75,60)(0,0)

\multiput(10,65)(70,0){2}{\vector(1,0){5}}
\multiput(10,55)(70,0){2}{\vector(1,0){5}}
\multiput(10,45)(70,0){2}{\vector(1,0){5}}
\multiput(10,15)(70,0){2}{\vector(1,0){5}}
\multiput(10, 5)(70,0){2}{\vector(1,0){5}}
\multiput(10,-5)(70,0){2}{\vector(1,0){5}}
%
% full lines
\put(5,65){\line(1,0){85}}
\put(5,55){\line(1,0){35}}
\put(5,45){\line(1,0){25}}
\put(5,15){\line(1,0){25}}
\put(5,5){\line(1,0){35}}
\put(5,-5){\line(1,0){85}}

\put(60,55){\line(1,0){30}}
\put(60, 5){\line(1,0){30}}

\put(70,45){\line(1,0){20}}
\put(70,15){\line(1,0){20}}

\put(30,15){\line(3,4){30}}
\put(30,45){\line(3,-4){30}}

\put(40, 5){\line(3,4){30}}
\put(40,55){\line(3,-4){30}}

% OGE between (25,15) and (25,45) (q and q1)
\put(25,15){\dashbox{2}(0,50){}}
\multiput(25,15)(0,50){2}{\circle*{2}}

\end{picture}
}

\end{picture}

%*******************************************************************
\vspace{1.0cm}
%*******************************************************************
%d 8

\setlength{\unitlength}{1.6pt}

\begin{picture}(320,70)(30,-5)
\put(70,28) {\makebox(0,0)[1]{ $D_8$ } }
\put(100,28) {\makebox(0,0)[1]{ = } }
\put(280,28) {\makebox(0,0)[1]{(16)} }
\put(220,-5) {\makebox(0,0)[1]{.} }

\put(120,0){

\begin{picture}(75,60)(0,0)

\multiput(10,65)(70,0){2}{\vector(1,0){5}}
\multiput(10,55)(70,0){2}{\vector(1,0){5}}
\multiput(10,45)(70,0){2}{\vector(1,0){5}}
\multiput(10,15)(70,0){2}{\vector(1,0){5}}
\multiput(10, 5)(70,0){2}{\vector(1,0){5}}
\multiput(10,-5)(70,0){2}{\vector(1,0){5}}
%
% full lines
\put(5,65){\line(1,0){85}}
\put(5,55){\line(1,0){35}}
\put(5,45){\line(1,0){25}}
\put(5,15){\line(1,0){25}}
\put(5,5){\line(1,0){35}}
\put(5,-5){\line(1,0){85}}

\put(60,55){\line(1,0){30}}
\put(60, 5){\line(1,0){30}}

\put(70,45){\line(1,0){20}}
\put(70,15){\line(1,0){20}}

\put(30,15){\line(3,4){30}}
\put(30,45){\line(3,-4){30}}

\put(40, 5){\line(3,4){30}}
\put(40,55){\line(3,-4){30}}

% OGE between (25,15) and (25,45) (q and q1)
\put(25,15){\dashbox{2}(0,30){}}
\multiput(25,15)(0,30){2}{\circle*{2}}

\end{picture}
}

\end{picture}

\setcounter{equation}{16}
%*******************************************************************
\vskip 1.0cm

In a given baryon-baryon scattering process the matrix element $h_{fi}$
in (\ref{eq:hfi})
is a weighted sum of the eight spatial overlap integrals represented
by the diagrams $D_1\dots D_8$,
\begin{equation}
h_{fi} = \sum_{n=1}^8 \; w_n\, I_n(D_n) \ .
\label{eq:hfi2}
\end{equation}
The weight of each diagram (introduced in \cite{BSKN})
is the product of a color factor, a fermion
permutation phase called the ``signature" of the diagram (which is $-1$ for
$D_1\dots D_4$ and $+1$ for $D_5\dots D_8$), the overall $(-)$ in $H_I$
(\ref{eq:HI}), and a reaction-dependent spin-flavor factor. The derivation of
these factors and the spatial overlap integrals they multiply is discussed in
detail elsewhere \cite{BS,BSKN}, so here we will simply present results with
minimal discussion. There is a minor change in the convention
for diagram weights relative to our earlier reference. In our KN study
\cite{BSKN} we incorporated the $(-)$ phase of $H_I$ and the signature phase in
the spatial overlap integral; here we include them in the diagram weight. This
overall factor was $(+1)$ for all KN diagrams, so the KN weights are unchanged
by our new convention.
Similarly, in our first paper \cite{BS} we incorporated the $(-)$ phase of
$H_I$ in the spatial overlap integral.
Our new convention is useful because it makes all the NN spatial overlap
integrals considered here positive,
so the overall amplitude phases are clear from the weights alone.

The color factors of the diagrams $D_1\dots D_8$ are
\begin{equation}
I_{color}([D_1\dots D_8]) \; = \;
\Big[\; +4/9, -2/9, -2/9, +1/9, +4/9, -2/9, -2/9, +1/9\; \Big] \ .
\end{equation}
$D_n$ is related to $D_{n+4}$ by $t\leftrightarrow u$ crossing; for
this reason the weights $w_n$ and $w_{n+4}$ are closely related in many
reactions.
To
simplify our presentation, when possible we will just give results for the
weights of $D_1\dots D_4$ and indicate the relative phase of the set for
$D_5\dots D_8$ after a bar. Thus for $\{ I_{color}(D_n) \} $ above we write
\begin{equation}
I_{color} \; = \;
\Big[ \; +4/9, -2/9, -2/9, +1/9\; \Big{|}\; (+)\; \Big] \ .
\end{equation}

The spin-flavor weights (incorporating the signature phases)
are just matrix elements of
the operator $\vec S_i \cdot \vec S_j$ between two initial quarks for the
given process. As an example, for $\Delta^{++}\Delta^{++}$, $S=3, S_z=3$
scattering there are four $|u_+u_+u_+\rangle /\sqrt{6}$ external baryons, and
only the $S^z_i S^z_j$ terms contribute. On summing over all scattering
diagrams in this channel we find
\begin{equation}
I_{spin-flavor}\cdot I_{signature}\Bigg{|}_{\Delta\Delta, I=3, S=3} \;  = \;
\Big[ \; +9/4, +9/2, +9/2, +9\; \Big{|}\; (-)\; \Big] \ ,
\end{equation}
which is $I_z$- and $S_z$-independent. Combining these we find the
$I=3, S=3 \ \Delta\Delta$ diagram weights,
\begin{equation}
\{ w_n ( \Delta\Delta, \; I=3, S=3 ) \} =
\Big[ \; +1, -1, -1, +1\; \Big{|}\; (-)\; \Big] \ .
\label{eq:wn}
\end{equation}
Since the spatial overlap integrals $I_5(D_5) \dots I_8(D_8)$ are
equal to the integrals $I_1(D_1) \dots I_4(D_4)$ after $t\leftrightarrow u$
crossing, the relative $(-)$ phase in the diagram weights (\ref{eq:wn})
insures that
the $I=3, S=3$ $\Delta\Delta$ scattering amplitude $h_{fi}$ is spatially
antisymmetric, as required for a totally antisymmetric
fermion-fermion scattering amplitude that is
symmetric in the remaining degrees of freedom (I and S). This antisymmetry is a
nontrivial check of our spin-flavor combinatorics, since it is only evident
after the sum over individual quark-gluon scattering diagrams is completed.

The diagram weights for all I,S channels of NN, N$\Delta$ and $\Delta\Delta$
elastic scattering are tabulated at the end of the paper; these
and the overlap integrals constitute our
central results.

The spatial overlap integrals associated with the diagrams may be determined
using the simple diagrammatic techniques presented in Appendix C of reference
\cite{BS}. In these integrals all momenta implicitly three-dimensional, and
have
an overall spin-spin coefficient $\kappa_{ss}$ of
\begin{equation}
\kappa_{ss} = {8\pi\alpha_s\over 3 m_q^2} {1\over (2\pi)^3} \ .
\end{equation}
This is $(-1)$ times the $\kappa$ of \cite{BS}, since we have chosen to include
the
$(-)$ phase of $H_I$ (\ref{eq:HI}) in the diagram weight factor, as
discussed above. The integrals are
\vskip 0.5cm

\begin{displaymath}
I_1 = \kappa_{ss}
\int \!\!\! \int da_1 da_2 \;
\Phi_A(a_1,a_2,A-a_1-a_2)
\Phi_C^*(a_1,a_2,C-a_1-a_2)
\end{displaymath}
\begin{equation}
\cdot
\int\!\!\! \int db_2 db_3 \;
\Phi_B(-A-b_2-b_3,b_2,b_3)
\Phi_D^*(-C-b_2-b_3,b_2,b_3) \ ;
\label{eq:i1}
\end{equation}
\begin{displaymath}
I_2 = \kappa_{ss}
\int\!\!\! \int\!\!\! \int\!\!\! \int da_1 da_3 db_2 dc_3 \;
\Phi_A(a_1,A-a_1-a_3,a_3)
\Phi_C^*(a_1,C-a_1-c_3,c_3)
\end{displaymath}
\begin{equation}
\Phi_B(C-A+a_3,b_2,-C-a_3-b_2)
\Phi_D^*(a_3,b_2,-C-a_3-b_2) \ ;
\label{eq:i2}
\end{equation}
\begin{displaymath}
I_3 = \kappa_{ss}
\int\!\!\! \int\!\!\! \int\!\!\! \int da_1 da_2 db_3 dd_1 \;
\Phi_A(a_1,a_2,A-a_1-a_2)
\Phi_C^*(a_1,a_2,C-a_1-a_2)
\end{displaymath}
\begin{equation}
\Phi_B(C-a_1-a_2,-A-C+a_1+a_2-b_3,b_3)
\Phi_D^*(d_1,-C-b_3-d_1,b_3) \ ;
\label{eq:i3}
\end{equation}
\begin{displaymath}
I_4 = \kappa_{ss}
\int\!\!\! \int\!\!\! \int\!\!\! \int da_1 da_3 db_1 db_3 \;
\Phi_A(a_1,A-a_1-a_3,a_3)
\Phi_C^*(a_1,C-a_1-b_1,b_1)
\end{displaymath}
\begin{equation}
\Phi_B(b_1,-A-b_1-b_3,b_3)
\Phi_D^*(a_3,-C-a_3-b_3,b_3) \ .
\label{eq:i4}
\end{equation}
We evaluate these in the c.m. frame, so the $t\leftrightarrow u$ crossed
integrals $I_5\dots I_8$ can be obtained by exchanging
$\vec C$ and $\vec D = -\vec C$, or in terms of the cosine of the
c.m. scattering angle
$\mu = \cos ( \theta_{\rm c.m.})$,
\begin{equation}
I_{n+4}(\mu ) =
I_{n}(-\mu ) \ .
\end{equation}
A simplification follows if all baryons have the same spatial wavefunctions,
as we assume here; in this case $I_2=I_3$ and hence $I_6=I_7$.

The overlap integrals may be carried out in closed form given
single-Gaussian wavefunctions (\ref{eq:wfn}), and each
gives a result of the form
\begin{equation}
I_n = \kappa_{ss}\, \eta_n \, \exp\bigg\{ -(A_n - B_n\mu ) P^2 \bigg\} \ ,
\label{eq:ints}
\end{equation}
where $P$ is the magnitude of the c.m. three-momentum of each baryon,
$P^2 = \vec A^2 = \vec B^2 = \vec C^2 = \vec D^2$.
The results are
\begin{equation}
I_1 = \kappa_{ss}
\exp\bigg\{ -{1\over 3\alpha^2} (\vec A - \vec C)^2 \bigg\} =
\kappa_{ss}
\exp\bigg\{ -{1\over 3\alpha^2} 2(1-\mu )P^2 \bigg\} =
\kappa_{ss}
\exp\bigg\{ {t\over 3\alpha^2} \bigg\} \ ;
\label{eq:i1g}
\end{equation}
\begin{displaymath}
I_2 = I_3 =
\kappa_{ss} \;
\Bigg( {12\over 11} \Bigg)^{3/2}
\exp\bigg\{  -{1\over 33\alpha^2} (20-12\mu )P^2 \bigg\}
\end{displaymath}
\begin{equation}
=
\kappa_{ss}\;
\Bigg( {12\over 11} \Bigg)^{3/2}
\exp\bigg\{ -{2(s-4M^2)\over 33\alpha^2} \bigg\}
\exp\bigg\{ {2t\over 11\alpha^2 } \bigg\} \ ;
\label{eq:i2g}
\end{equation}
\begin{equation}
I_4 =
\kappa_{ss}\;
\Bigg( {3\over 4} \Bigg)^{3/2}
\exp\bigg\{  -{1\over 3\alpha^2} P^2 \bigg\} =
\kappa_{ss}\;
\Bigg( {3\over 4} \Bigg)^{3/2}
\exp\bigg\{ -{(s-4M^2)\over 12\alpha^2} \bigg\} \ .
\label{eq:i4g}
\end{equation}
In the
final expression for each integral
we have substituted for $P^2$ and $\mu$ in terms of the Mandelstam
variables $s$ and $t$ using relativistic kinematics, $s=4(P^2+M^2)$ and
$t=-2(1-\mu)P^2$.

Near threshold the overlap integrals are comparable in magnitude, but at higher
energies their behaviors differ markedly. All but $I_1$ and its
crossing-symmetric partner $I_5$
are strongly suppressed in $s$; the diagrams
$D_1$ and $D_5$ therefore dominate at high energies, for forward and backward
scattering respectively.
This behavior is due to
the mechanism of ``minimum spectator suppression",
as was discussed in detail
in section IIe of reference \cite{BSKN}. To summarize the arguments for
this case:
1) for forward scattering, $\vec A = \vec C$, diagram $D_1$ (9)
requires no spectator to cross into
an opposite-momentum hadron, which would carry considerable
suppression due to a small wavefunction overlap;
only the hard-scattered constituents are required
to reverse momentum.
2) $D_5$ (13) requires all spectators that were initially in a baryon
with momentum $\vec A$ to reside finally in a baryon with momentum $\vec D$.
Clearly the suppression due to wavefunction overlaps
of the spectators will be less important
if the final baryon D
has the same momentum as the initial baryon A,
$\vec D = \vec A$. This corresponds to backscatter,
$\vec C = -\vec A$, since $\vec C = -\vec D$ in the c.m. frame.
As in $D_1$, only the hard-scattered quarks are then required
to recoil into a baryon with three-momentum opposite to that of their
initial baryon. These two explanations are actually equivalent because
$D_1$ and $D_5$ are related by crossing.

\section{NN core potentials and phase shifts}

The Hamiltonian matrix elements
\begin{equation}
h_{fi} = \sum_{n=1}^8 \ w_n I_n
\end{equation}
for the four I,S channels accessible in NN scattering are summarised
by the diagram weights in Table I. Specialising to the even-L channels
I,S=0,1 and 1,0, for which a repulsive core in S-wave
is a well known feature,
we see that all eight coefficients $\{ w_1 \dots w_8 \} $
are positive or zero in both cases, corresponding to a
repulsive interaction. For a more quantitative evaluation, we
can relate this $h_{fi}$ matrix element
to an NN potential near threshold, which is defined
to give the same low-energy scattering amplitude near threshold
in Born approximation. (See Appendix E of reference \cite{BS} for a detailed
discussion.)
For an $h_{fi}$ of the form
\begin{equation}
h_{fi} = {8\pi \alpha_s \over 3 m_q^2} {1\over (2\pi)^3}
\sum_{n=1}^4 \; w_n \eta_n \exp\bigg\{ -(A_n - B_n\mu ) P^2 \bigg\}
\end{equation}
the Born-equivalent potential is
\begin{equation}
V_{NN}(r) = {8 \alpha_s \over 3 \sqrt{\pi} m_q^2}
\sum_{n=1}^4 \; { w_n \eta_n \over (A_n + B_n )^{3/2} }
\exp\bigg\{ -{r^2\over (A_n + B_n)}  \bigg\} \ .
\label{eq:VNN}
\end{equation}
The $t\leftrightarrow u$ crossed
diagrams $D_5\dots D_8$ are not included in this sum because they will
automatically be generated by the crossed diagram in NN$\to$NN potential
scattering through $V_{NN}(r)$.

The numerical potentials predicted for S-wave I=0 and I=1 NN systems
are shown in Fig.1
for our ``reference" set of quark model parameters \cite{BSKN},
$\alpha_s=0.6$, $m_q=0.33$ GeV and
$\alpha=0.4$ GeV.
Actually only the two parameters $\alpha_s/m_q^2=5.51$ GeV$^{-2}$ and
$\alpha=0.4$ GeV are involved in $V_{NN}(r)$.
These potentials are
consistent with expectations for NN core interactions; they are
repulsive and have ranges of about 1/2 fm and peak values
comparable to +1 GeV, which is essentially infinite from a nuclear physics
viewpoint. It may be interesting in future work to parametrize the amplitudes
associated with each diagram (the weights in Table I) as a two-nucleon spin-
and isospin-interaction of the form $A\, I + B\, \vec S_1\cdot \vec S_2 +
C\, \vec\tau_1\cdot\vec\tau_2 + D\, \vec S_1\cdot \vec S_2\,
\vec\tau_1\cdot\vec\tau_2$,
which will allow a more direct comparison with meson-exchange models
\cite{Gross}.

Low-energy equivalent NN core potentials have been presented as the results of
some of the NN resonating-group and variational calculations we discussed in
the introduction.
In their Figs.1 and 2
Suzuki and Hecht \cite{SH} show numerical results for the
NN core potentials of Harvey \cite{Harvey},
Faessler, Fernandez, L\"ubeck and Shimizu \cite{FFLS} and
Oka and Yazaki \cite{OY}.
The Faessler {\it et al.} and Oka-Yazaki potentials are quite similar to our
potentials in Fig.1, with values at the origin between 0.6 and 1.0 GeV and
comparable ranges. Harvey finds potentials with somewhat longer ranges,
which Suzuki and Hecht attribute to
his choice of a larger nucleon width parameter, $b_N\equiv 1/\alpha=0.8$ fm;
the Faessler {\it et al.} and Oka-Yazaki values are 0.475 fm
and 0.6 fm respectively, and we use a comparable $1/\alpha=0.493$ fm.

The NN core potentials found by Maltman and Isgur, in Fig.1 of reference
\cite{MI}, also have similar ranges but are somewhat larger in magnitude,
V(0)=1.2 GeV for I=0 and 2.3 GeV for I=1. Our difference in the contact values
is due in part to the choice of parameters;
in our calculations the potentials
are proportional to $\alpha_s \alpha^3 /m_q^2$, which is 0.353 GeV with our
parameters and 0.489 GeV for Maltman and Isgur.
Note, however, that no other
reference finds the large Maltman-Isgur I=1/I=0 ratio at contact. The value
chosen for $\alpha_s \alpha^3 /m_q^2$ by itself does not explain the
differences between potentials; those reviewed by Suzuki and Hecht use
$\alpha_s \alpha^3 /m_q^2 \approx 0.55$ GeV,
so we would naively expect our potentials to be $\approx 0.6$ times
as large as theirs. Of course the values near the origin have little physical
relevance due to their small Jacobean weight, and in any case
we are comparing potentials derived using three
different methods, and these differences may preclude a more accurate
comparison of results.

The choice of parameters will be discussed in more detail in the section on
differential cross sections. Here we simply note that the
smaller value of $\alpha_s$ we use
is now generally
preferred because recent spectroscopy studies using improved
wavefunctions have considerably lowered the value required to fit
hadron spectroscopy.
The NN references we compare with predate the improved
spectroscopy studies and thus used a rather large value of $\alpha_s$, which
was required to give a realistic N$\Delta$ splitting given single-Gaussian
wavefunctions.
We also prefer to use our fixed parameter set because these
values were found to give reasonable results for low-energy S-wave
$\pi\pi$, K$\pi$ and KN scattering in our previous studies \cite{BS,BSW,BSKN}.

Note in Fig.1 that the intermediate-range attractions which are responsible for
the deuteron in I=0 and its almost-bound I=1 partner are absent from our
quark Born potentials. This is as we anticipated, given that these attractions
arise mainly from a spatial distortion of interacting-nucleon wavefunctions
\cite{Isgur,MI}; in our leading-order Born calculation we assume fixed
nucleon spatial wavefunctions. The attraction presumably
arises at higher order in the Born series, and may be accessible through
leading-order Born calculations of off-diagonal matrix elements.

Oka and Yazaki (Fig.2 of \cite{OY}) and Koike (Fig.3 of \cite{Koike}) also show
the S-wave phase shifts which result from their NN core interactions. In Fig.2
we show the S-wave phase shifts we find on numerically integrating
the Schr\"odinger equation with the potentials of Fig.1. Our phase shifts are
very similar to the results of these earlier resonating-group studies.
Although we would like to compare our phase shifts to experiment directly,
the
experimental phase shifts \cite{VPI} are unfortunately complicated by the
presence of the deuteron and its I=1 partner near
threshold. These states will have to be incorporated in our calculation before
we can make a useful comparison between our theoretical core phase shifts and
experiment.

The Born-order approximate phase shifts
(proportional to $h_{fi}$) can be determined analytically
using Eq.(6) of \cite{BSW} and dividing by 2 for identical particles,
\begin{equation}
\delta^{(\ell )} =
-{\pi^2 \over 2} P E_P\; \int_{-1}^1 \; h_{fi}(\mu )P_{\ell}(\mu) .
\label{eq:phasesh}
\end{equation}
The momentum, energy and $\mu = \cos (\theta_{\rm c.m.} )$
are for one nucleon in the c.m. frame.
{}From our general result for $h_{fi}$ (\ref{eq:hfi2}) and the
Gaussian-wavefunction integrals (\ref{eq:ints}),
we find an $\ell$th Born-order partial-wave phase shift of
\begin{equation}
\delta^{(\ell )} =
-{\alpha_s \over 3 m_q^2} P E_P\sum_{n=1}^8\; w_n \eta_n e^{-A_n P^2}
i_{\ell}(B_n P^2) \ .
\label{eq:psdl}
\end{equation}
The spin-dependence of this result is implicit in the weights $\{ w_n \}$.
Note that these phase shifts are functions of $\ell$ and spins only,
so there is no spin-orbit force in our effective NN interaction.
This is as expected given that our only interaction at the quark level is the
spin-spin hyperfine term. A more realistic model will require a generalization
to include the OGE spin-orbit term and perhaps coupled channel effects, as
we discussed in our study of KN scattering.

The analytic Born-order result (\ref{eq:psdl}) for the phase shifts is
unfortunately of little
utility for NN S-waves given realistic quark-model parameters;
the equivalent potential $V_{NN}(r)$ is nonperturbatively large in this
case, and must be iterated coherently to determine phase shifts,
as we have done in Fig.2 using the NN Schr\"odinger equation.
In contrast to this result, we previously found
nonperturbative effects in the S-wave phase shifts of
$\pi\pi$, KK and KN
systems to be much less important, due to somewhat shorter-ranged
forces and the smaller reduced mass.
The NN Born-approximation phase shifts
(\ref{eq:psdl}) are presumably more useful for
higher partial waves and higher energies, since multiple scattering
effects are expected to be largest in S-wave near threshold.

\vskip 0.5cm

\section{other nonstrange BB$'$ channels: N$\Delta$ and $\Delta\Delta$}

The core potentials predicted by the quark Born formalism for other nonstrange
baryon-baryon channels should allow tests of the assumed hyperfine dominance in
systems other than the familiar S-wave NN cases. The short-range interactions
in the N$\Delta$ and $\Delta\Delta$ channels may be observable experimentally
as final state interactions or, if the interaction is sufficiently strong to
support bound states, as dibaryon molecules not far below threshold. The
possibility of nonstrange resonances in the $q^6$ sector has been considered by
many authors, the earliest reference apparently being a group-theoretic study
by Dyson and Xuong \cite{DX}. These $q^6$ systems have also been studied using
the bag model (which is unfortunately known to give unphysical predictions of a
host of multiquark resonances), one-boson-exchange models and the
nonrelativistic quark model; references before 1985 are summarised by Maltman
\cite{Maltman}.

Our results for N$\Delta$ and $\Delta\Delta$ are summarised by the diagram
weights in Tables II and III. After completing their derivation we found that
some of these matrix elements had previously been tabulated by Suzuki and Hecht
\cite{SH}; our NN and $\Delta\Delta$ weights $w_1, w_2, w_3, w_4$ are
equivalent to the coefficients $C_{ST}^{(5)}, -C_{ST}^{(2)}, -C_{ST}^{(3)},
C_{ST}^{(4)}$ in their Table II, which provides an independent check of our
results in these cases. As these weights multiply comparable spatial overlap
integrals which give positive contributions to the low-energy
equivalent baryon-baryon potential ({\ref{eq:VNN}), negative weights imply
attractive potential contributions.

Referring to Table II, we see that the N$\Delta$ system has a strongly
repulsive core in the channels I,S=2,1 and 1,2 and weak core interactions in
2,2 and 1,1. Unlike the $\Delta\Delta$ system (to be discussed subsequently)
our N$\Delta$ core interactions do not lead to bound states in any channel.
Since the lightest reported dibaryons have masses very close to the N$\Delta$
threshold \cite{VPI,Seth,PDG86},
the experiments may be seeing threshold effects due
to the opening of the N$\Delta$ channel, or perhaps weakly-bound N$\Delta$
molecules. Our calculation does not support the existence of such bound states,
although the NN system is similarly predicted to have a purely repulsive core,
but the I,S=0,1 deuteron is nonetheless bound by an intermediate-range
attraction which is absent from our leading-order Born calculation. Similar
weakly-bound states may exist in N$\Delta$ and $\Delta\Delta$ as well, despite
repulsive cores.

Next we consider the $\Delta\Delta$ system.
For $\Delta\Delta$ some general rules follow from the assumption of
a single $\lambda \cdot \lambda$ interaction; since the initial
three-quark clusters are transformed into color octets by the interaction,
line diagrams with zero
or three pairs of quarks exchanged are forbidden. Thus the amplitude for
$\Delta^{++}\Delta^-$ elastic scattering must be zero. This
implies relations between $\Delta\Delta$ amplitudes with different isospins,
\begin{equation}
h_{fi}(\Delta\Delta; I=0,S) =  - h_{fi}(\Delta\Delta; I=2,S)
\end{equation}
and
\begin{equation}
h_{fi}(\Delta\Delta; I=1,S) =  -{1\over 9}\, h_{fi}(\Delta\Delta; I=3,S)  \ .
\end{equation}
Specializing to the $(+)$-symmetry (even-L) cases in Table III,
which include the S-wave
channels that are {\it a priori} the most likely to support bound states, it
is evident that two $\Delta\Delta$
channels have strongly attractive core potentials, I,S=1,0 and 0,1. Of
these 0,1 has the strongest attraction. We search for
bound states by solving the Schr\"{o}dinger equation in the
$\Delta\Delta$ system using the low-energy potential (\ref{eq:VNN}). With our
reference parameter set
$\alpha_s/m_q^2 = 5.51$ GeV$^{-2}$ and $\alpha = 0.4$ GeV
the
attractive core is too weak to induce binding.
Note, however, that previous studies of baryons using single-Gaussian
wavefunctions have generally assumed a much
stronger hyperfine term, for reasons we will
discuss subsequently. If we use a typical parameter set from these references,
$\alpha_s/m_q^2 = 14.9$ GeV$^{-2}$
and $\alpha = 0.32$ GeV
(Maltman and Isgur \cite{MI}),
we find a single S-wave $\Delta\Delta$ bound state in the 0,1 channel, with
$E_B=40$ MeV. Of course this channel has a fall-apart coupling to
NN, so a coupled-channel treatment including the NN system may be required to
search for resonant effects. None of the other $\Delta\Delta$ channels have
sufficiently strong attractive cores to form bound states in our formalism
with the Maltman-Isgur parameters.

Our result for the I,S=0,1 $\Delta\Delta$ channel is remarkably similar to the
conclusion of Maltman \cite{Maltman}, who found that the 0,1 channel has the
strongest diagonal attraction in the $\Delta\Delta$ Hilbert space, and that
these diagonal forces led to a $\Delta\Delta$ bound state with $E_B=30$ MeV.
Maltman concluded, however, that off-diagonal effects due to the excitation of
hidden-color states eliminated this bound state and led to binding in I,S=3,0
($E_B=30$ MeV) and 0,3 ($E_B=260$ MeV) instead. The 3,0 and 0,3 channels had
previously been suggested as possibilities for $\Delta\Delta$ bound states
\cite{OY,KF,CGMR}. In contrast we find strong repulsion in the 3,0 channel
and a weak core in 0,3.

In view of the parameter- and approximation-dependence of predictions of
$\Delta\Delta$ bound states and the theoretical uncertainties in treating
hidden-color basis states, the possibility of nonstrange dibaryon molecules
should be regarded as an open question for experimental investigation. The
channels which appear of greatest interest at present are the attractive-core
systems I,S=0,1 and 1,0 and the 3,0 and 0,3 channels, which previous studies
suggested as possibilities for binding.

\section{NN differential cross sections}

We can use Eq.28 of reference \cite{BS},
\begin{equation}
{d\sigma \over dt} = {4\pi^5 s\over (s-4M^2)} |h_{fi}|^2 \ ,
\label{eq:dsig1}
\end{equation}
to determine the nucleon-nucleon differential cross section in leading
Born approximation, given the NN $h_{fi}$ matrix element (\ref{eq:hfi2}).
For the experimentally well-determined case of unpolarised PP elastic
scattering, we have a weighted sum of the S=0 and S=1 differential cross
sections,
\begin{equation}
{d\sigma \over dt}\bigg{|}_{\rm PP, unpolarised} =
{1\over 4} {d\sigma \over dt}\bigg{|}_{I=1, S=0} +
{3\over 4} {d\sigma \over dt}\bigg{|}_{I=1, S=1} \ .
\label{eq:dsig2}
\end{equation}
To obtain the S=0 and S=1 cross sections one simple substitutes the
appropriate I=1 diagram weights $\{ w_n \}$ from Table I and the integrals
$\{ I_n\} $ from (\ref{eq:i1g}-\ref{eq:i4g}).

Before we discuss our prediction for this differential cross section we briefly
recall the experimental unpolarised PP result. This is shown for a range of
$P_{\rm lab}$ in Fig.3, adapted from
Ryan {\it et al.} \cite{Ryan},
Ankenbrandt {\it et al.} \cite{Ank},
Clyde {\it et al.} \cite{Clyde},
Allaby {\it et al.} \cite{Allaby}, and from the
ISR data of Nagy {\it et al.} \cite{Nagy} and Breakstone
{\it et al.} \cite{Breakstone}. The data in the figure were
obtained from the Durham-Rutherford HEP data archive.
Near threshold the angular distribution is approximately isotropic, but as
$P_{\rm lab}$
increases the scattering at large angles falls rapidly, and at high
energies the differential cross section is dominated by an asymptotic
``diffractive peak",
\begin{equation}
\lim_{s\to\infty, \  |t|/s<<1} \ \ \ {d\sigma \over dt}^{\rm expt.} \approx
a  e^{bt} \ .
\label{eq:asymp}
\end{equation}
For the purely hadronic part (as distinct from the divergent
forward Coulomb peak)
one finds
\begin{equation}
a^{\rm expt.} \approx 70 \; \hbox{mb GeV}^{-2}
\end{equation}
and
\begin{equation}
b^{\rm expt.} \approx 11-12 \; \hbox{GeV}^{-2}
\end{equation}
for the asymptotic form \cite{Breakstone}.

On evaluating (\ref{eq:dsig2}) for NN scattering using (\ref{eq:hfi2}),
(\ref{eq:dsig1}) and Table I, we find that several features of the experimental
differential cross section are successfully reproduced by our Born-order
calculation. The theoretical Born-order cross section (\ref{eq:dsig2}) which
follows from our ``reference" parameter set $\alpha_s/m_q^2=5.51$ GeV$^{-2}$
and $\alpha=0.4$ GeV is shown in Fig.4 for $P_{\rm lab}=1.05, 1.75, 3.$
and $10.$
GeV, selected for comparison with Fig.3. Although $P_{\rm lab}=10.$ GeV
superficially appears to be very relativistic, in the c.m. frame it actually
corresponds to $P_{\rm c.m.}=2.07$ GeV, which for nucleon-nucleon scattering is
only quasirelativistic.

First note that the smooth evolution from an isotropic angular distribution to
an asymptotic forward-peaked one with increasing $s$ is a simple consequence of
the suppression in $s$ of all diagrams at small $|t|$
except $D_1$. In our calculation the
contributions of the other diagrams fall exponentially with $s$. The
experimental large-angle scattering does not fall this rapidly, and the
discrepancy is probably due to our use of single-Gaussian forms; the actual
proton wavefunction has short-distance
quark-quark correlations, which presumably
lead to power-law contributions at large $s$ and $|t|$.

Second, the observed approximate asymptotic form (\ref{eq:asymp}) is actually
predicted by our single-Gaussian Born calculation. The overall normalization
$a$ and slope parameter $b$ for this process are predicted to be
\begin{equation}
a = {4\pi \alpha_s^2\over 9 m_q^4}
\Bigg( \;
{1\over 4}\, (w_1^{I=1,S=0})^2 +
{3\over 4}\, (w_1^{I=1,S=1})^2
\Bigg)  =
{6364 \over 19683}  {\pi \alpha_s^2\over m_q^4}
\label{eq:a}
\end{equation}
and
\begin{equation}
b = {2\over 3\alpha^2} \ .
\end{equation}
The theoretical result (\ref{eq:a}) for the magnitude of the forward
peak is actually independent of the spatial wavefunction, since the defining
integral (\ref{eq:i1}) is just the product of two normalization integrals in
the limit $\vec A = \vec C$.

With the reference parameter set
we predict a somewhat smaller, broader peak than is observed experimentally,
with
\begin{equation}
a = 12. \; \hbox{mb GeV}^{-2}
\end{equation}
and
\begin{equation}
b = 4.2 \; \hbox{GeV}^{-2} \ .
\end{equation}
Both $a$ and $b$, however, are sensitive to the choice of quark model
parameters, and vary by factors of about 10 and 3 respectively when
$\alpha_s/m_q^2$ and $\alpha$ are varied through a plausible range,
which we shall discuss below.
If we use typical ISR experimental intercept and slope values of
$a=70.$ mb GeV$^{-2}$ and $b=11.$ GeV$^{-2}$ \cite{Breakstone} as input
to fix our two parameters, the fitted values are
\begin{equation}
{\alpha_s\over m_q^2} \; = 13.3 \; \hbox{GeV}^{-2}
\label{eq:hfstr}
\end{equation}
corresponding to $m_q=0.21$ GeV if we leave $\alpha_s=0.6$, and
\begin{equation}
\alpha = 0.246 \; \hbox{GeV} \ .
\end{equation}
Although these fitted
parameters give the observed intercept $a$ and slope $b$, the higher-$|t|$
wings of the resulting distribution fall much too rapidly with $s$. This is
prob
ably
an artifact of our use of soft single-Gaussian wavefunctions, and a
calculation of the differential cross sections which follow from
more realistic wavefunctions will be a very interesting exercise.

The fitted strength of the hyperfine interaction (\ref{eq:hfstr})
is now believed to be rather
large, although it is similar to the values used in many
previous quark-model studies of baryons. Examples of previous values in
chronological order are $\alpha_s / m_q^2  = 15.5$   GeV$^{-2}$ (Oka and
Yazaki, 1980 \cite{OY}); $37.8$ GeV$^{-2}$ (Harvey, 1981 \cite{Harvey}; this
value now appears exceptionally large); $7.7$ GeV$^{-2}$ (Faessler {\it et
al.}, 1982 \cite{FFLS}); $14.9$ GeV$^{-2}$ (Maltman and Isgur, 1984 \cite{MI});
and $14.4$ GeV$^{-2}$ (Koike, 1986 \cite{Koike}). Large values were
required to fit the N-$\Delta$ mass splitting given single-Gaussian
wavefunctions; since this is proportional to $(\alpha_s/m^2_q)|\psi(0)|^2$, an
underestimated wavefunction at contact must be compensated for by a large
$(\alpha_s/m^2_q)$. If one instead uses the actual Coulomb plus linear
wavefunctions from the nonrelativistic Schr\"odinger equation with $m_q\approx
0.3$ GeV, the larger value of $|\psi(0)|$ leads to a much smaller
$\alpha_s/m^2_q\approx 5$ GeV$^{-2}$. Since our scattering calculation uses
Gaussian wavefunctions, one could argue which parameter value is most
appropriate; an improved calculation with more realistic wavefunctions will
probably be required to eliminate these parameter uncertainties.

The baryon width parameter $\alpha$ has also been assigned a rather large range
of values in previous work. Representative values in chronological order
are
$\alpha =
 0.41$ GeV (Copley, Karl and Obryk, baryon photocouplings, 1969 \cite{CKO});
$0.32$ GeV (Isgur and Karl, baryon spectroscopy, 1979\cite{IK});
$0.41$ GeV (Koniuk and Isgur, baryon photocouplings, 1980 \cite{KI});
$0.33$ GeV (Oka and Yazaki, NN interactions, 1980 \cite{OY});
$0.25$ GeV (Harvey, NN interactions, 1981 \cite{Harvey});
$0.42$ GeV (Faessler {\it et al.}, NN interactions, 1982 \cite{FFLS});
$0.25$ GeV (Hayne and Isgur relativised quark model, 1982 \cite{HI});
$0.32$ GeV (Maltman and Isgur, NN interactions, 1984 \cite{MI});
$0.34$ GeV (Koike, NN interactions, 1986 \cite{Koike});
$0.3$ and $0.42$ GeV, with the smaller
value preferred (Li and Close, baryon electroproduction, 1990 \cite{LC}).

Most potential models assume a value of $m_q$ near 0.3 GeV, although
the relativised models of Hayne-Isgur \cite{HI}, Godfrey-Isgur \cite{GI}
(mesons) and Capstick-Isgur (baryons) \cite{CI}
use a lower value of
0.22 GeV. Although these relativised models also use a small
value of $\alpha_s(Q^2=0)=0.6$
for the infrared limit of an effective running $\alpha_s(Q^2)$,
the hyperfine strength $\alpha_s(0)/m_q^2 = 12.2$ GeV$^{-2}$
is again large because of the smaller $m_q$.
Its effects are
reduced, however, by the
use of a ``smeared" contact interaction.

The $m_q$ and $\alpha$ used by Hayne and Isgur are
essentially identical to the values we need to fit the
slope and intercept of the experimental PP diffractive peak,
although this is presumably fortuitous agreement.

Although we can fit the magnitude and $t$-dependence of the small-$|t|$
differential cross section at high energies reasonably well with our quark Born
results, we emphasize that this is at best an incomplete description of
diffractive scattering, because the Born amplitude is purely real whereas the
experimental small-$|t|$ amplitude is known to be close to imaginary
\cite{Nagy,UA4}. This may imply that the first Born approximation is inadequate
for small-$|t|$ diffractive scattering, and that the coupling to inelastic
channels is an essential component of a description of the diffractive
amplitude, even for elastic processes \cite{Roger}. It may be necessary to
iterate the effect of diagram $D_1$ to generate the observed phase
\cite{Roberts}, perhaps
including a sum over virtual inelastic channels. Our conclusion
that suppression due to the spectator lines is the dominant origin of the
observed diffractive $t$-dependence \cite{BSKN} would presumably be unchanged
by iteration of the $qq$ hard scattering process in diagram $D_1$.

In view of the complexity of the baryon-baryon scattering problem, which
involves a sum of thousands of diagrams and the evaluation of 36-dimensional
overlap integrals, and the questionable accuracy of our nonrelativistic
single-Gaussian wavefunctions, we find our approximate agreement with
experiment encouraging. The most important discrepancies are in the phase of
the scattering amplitude (which may require a higher-order Born study) and in
the higher-$|t|$ ``wings" of the distribution, which may be more accurately
described by more realistic Coulomb plus linear baryon wavefunctions. A
determination of the proton-proton differential cross section given more
realistic nucleon wavefunctions would be a very interesting future application
of this formalism.

\section{Summary and conclusions}

We have applied the quark Born diagram formalism to nonstrange baryon-baryon
elastic scattering. In this approach the hadron-hadron scattering
amplitude is taken to be the sum of all
single
quark-pair
interactions followed by all allowed
quark interchanges, with nonrelativistic quark model wavefunctions
attached to the external lines.
This may be a useful description of
reactions which are free of $q\bar q$ annihilation. The model has few
parameters (here only two, the baryon oscillator parameter
$\alpha$ and the hyperfine strength $\alpha_s/m_q^2$, since we incorporate only
the OGE spin-spin hyperfine term in this study),
and with
Gaussian wavefunctions and a contact interaction
the scattering amplitudes can be derived analytically.
The model was previously applied to I=2 $\pi\pi$ and I=3/2 K$\pi$ scattering
with good results, and also gives reasonable results for low-energy S-wave
KN scattering,
although there are discrepancies at higher energies and in
higher partial waves.

NN scattering is an important test of this approach because it is also
annihilation-free (at the valence quark level), and the baryon
wavefunction and the dominance of the spin-spin OGE hyperfine
interaction in NN
are already reasonably
well established.
We find that the quark Born diagrams predict repulsive core interactions in
both S-wave NN channels, and the equivalent low-energy potentials we
extract from the scattering amplitudes are very similar to the results
of previous resonating-group and variational calculations. We also give results
for the N$\Delta$ and $\Delta\Delta$ core interactions induced by the OGE
spin-spin term, and find that certain $\Delta\Delta$ channels have attractive
cores and may possess bound states. Finally we determine the NN differential
cross section predicted by our Born amplitude, and compare the results with
the experimental unpolarised PP differential cross section over a wide range of
energies. We find that several well known features of experimental PP
scattering are evident in our Born results,
including the development of a high-energy forward peak with an
approximately correct width and magnitude.

\acknowledgements

We acknowledge useful contributions from W.Bugg, F.E.Close, G.Farrar,
H.Feshbach, F.Gross,
M.Guidry, N.Isgur, L.S.Kisslinger, Y.Koike, Z.Li, K.Maltman, R.J.N.Phillips,
G.R.Satchler, K.K.Seth, G.Sterman, M.R.Strayer and R.L.Workman.
This work was sponsored in part by the United States Department of Energy under
contracts DE-AC02-76ER03069 with the Center for Theoretical Physics at the
Massachusetts Institute of Technology and DE-AC05-840R21400 with Martin
Marietta Energy Systems Inc, and DE-AC05-84ER40150 with SURA / CEBAF.

\newpage

\begin{table}
Table I. Diagram weights for NN elastic scattering.\\
\begin{tabular}{rrrrrc}
&$\omega_1$&$\omega_2$&$\omega_3$&$\omega_4$&$\mbox{rel.  phase}\ \omega_5
\dots \omega_8$\\ \tableline I=1;\ S=1&$59 \over 81$&$17 \over 81$&$17 \over
81$&$10 \over 81$&$(-)$ \\ &&&&& \\     \ S=0&$31 \over 27$&$7 \over 27$&$7
\over 27$&$0$&$(+)$ \\ \tableline \tableline I=0;\ S=1&$19\over 27$&$7 \over
27$&$7 \over 27$&$2\over 27$&$(+)$ \\ &&&&& \\     \ S=0&$-{1\over 9}$&$5 \over
9$&$5 \over 9$&$0$&$(-)$
\end{tabular}
\end{table}

\vspace{1cm}
\begin{table}
Table II.  Diagram weights for N$\Delta $ elastic scattering.\\
\begin{tabular}{rrrrrrrrr}
&$\omega_1$&$\omega_2$&$\omega_3$&$\omega_4$&$\omega_5$&$\omega_6$&
$\omega_7$&$\omega_8$
\\ \tableline I=2;\ S=2&$7 \over 9$&$1 \over 9$&$-{5\over 9}$&$0$&$2 \over
9$&$2\over 9$&$-{4\over 9}$&$-{1\over 3}$ \\ &&&&&&&& \\      \ S=1&$37 \over
27$&$7 \over 27$&$1 \over 27$&$-{4 \over 27}$&$ -{2 \over 27}$&$ -{2 \over
27}$&$4 \over 27$&$-{1 \over 27}$ \\ \tableline \tableline I=1;\ S=2&$13 \over
27$&$7 \over 27$&$1\over 27$&$0$&$-{2\over 27}$&$-{2\over 27}$&$4 \over
27$&$1\over 9$ \\ &&&&&&&& \\      \ S=1&$7\over 81$&$49\over 81$&$-{29
\over 81}$&$-{28\over 81}$&$2\over 81$&$2\over 81$&$-{4 \over 81}$&$1\over
81$ \\
\end{tabular}
\end{table}

\vspace{1cm}

\begin{table}
Table III.  Diagram weights for $\Delta \Delta$ elastic scattering.
\begin{tabular}{rrrrrc}
&$\omega_1$&$\omega_2$&$\omega_3$&$\omega_4$&$\mbox{rel.  phase}\ \omega_5
\dots  \omega_8$\\ \tableline I=3;\ S=3&$1$&$-1$&$-1$&$1$&$(-)$\\ &&&&& \\
\ S=2&$5 \over 3$&$-{1 \over 3}$&$-{1\over 3}$&$-1$&$(+)$ \\ &&&&& \\     \
S=1&$19 \over 9$&$1 \over 9$&$1 \over 9$&$-{1 \over 9}$&$(-)$ \\ &&&&& \\     \
S=0&$ 7 \over 3$&$1 \over 3$&$1 \over 3$&$1$&$(+)$ \\ \tableline \tableline
I=2;\ S=3&$1 \over 3$&$-{1 \over 3}$&$-{1 \over 3}$&$1 \over 3$&$(+)$ \\ &&&&&
\\     \ S=2&$5 \over 9$&$-{1 \over 9}$&$-{1\over 9}$&$-{1 \over 3}$&$(-)$ \\
&&&&& \\     \ S=1&$19 \over 27$&$1 \over 27$&$1 \over 27$&$-{1 \over
27}$&$(+)$ \\ &&&&& \\     \ S=0&$7 \over 9$&$1 \over 9$&$1 \over 9$&$1\over
3$&$(-)$ \\ \tableline \tableline I=1;\ S=3&$-{1 \over 9}$&$1 \over 9$&$ 1
\over 9$&$-{1 \over 9}$&$(-)$ \\ &&&&& \\     \ S=2&$-{5 \over 27}$&$1 \over
27$&$1\over 27$&$1\over 9$&$(+)$ \\ &&&&& \\     \ S=1&$-{19 \over 81}$&$-{1
\over 81}$&$-{1 \over 81}$&$1 \over 81$&$(-)$ \\ &&&&& \\     \ S=0&$-{7 \over
27}$&$-{1\over 27}$&$-{1 \over 27}$&$-{1 \over 9}$&$(+)$ \\ \tableline
\tableline I=0;\ S=3&$-{1 \over 3}$&$1 \over 3$&$1 \over 3$&$-{1 \over
3}$&$(+)$ \\ &&&&& \\     \ S=2&$-{5 \over 9}$&$1 \over 9$&$1 \over 9$&$1 \over
3$&$(-)$ \\ &&&&& \\     \ S=1&$-{19 \over 27}$&$-{1 \over 27}$&$-{1 \over
27}$&$1 \over 27$&$(+)$ \\ &&&&&  \\     \ S=0&$-{7 \over 9}$&$-{1 \over
9}$&$-{1 \over 9}$&$-{1 \over 3}$&$(-)$

\end{tabular}

\end{table}


\begin{references}

\bibitem[1]{earlycore}
P.M.Morse, J.B.Fisk and L.I.Schiff, Phys. Rev. 50, 748 (1936);
H.A.Bethe and R.F.Bacher, Rev. Mod. Phys. 8, 82 (1936), acknowledge the
possibility of a potential of this general form
but were skeptical, see especially
Section 27, p.161;
subsequent early
references include G.Parzen and L.I.Schiff, Phys. Rev. 74, 1564 (1948);
R.Jastrow, Phys. Rev 79, 389 (1950) and references cited therein; {\it ibid.},
Phys. Rev 81, 165 (1951).

\bibitem[2]{Yukawa} H.Yukawa, Proc. Phys.-Mat. Soc. Japan 17, 48 (1935).

\bibitem[3]{OBE} See for example M.Lacombe, B.Loiseau, J.M.Richard, and
R.VinhMau, Phys. Rev. C21, 861 (1980);
K.Holinde, Phys. Lett. C68, 121 (1981);
R.Machleidt, Adv. Nucl. Phys. 19, 189 (1989);
E. Hummel and J.A. Tjon, Phys. Rev. C42, 423 (1990);
F.Gross, J.W.VanOrden and K.Holinde, Phys. Rev. C45, 2094 (1992).

\bibitem[4]{Isgur}
N.Isgur, Acta Physica Austriaca,
Suppl. XXVII, 177-266
(1985).

\bibitem[5]{siemens} P.J.Siemens and A.S.Jensen, {\it Elements of Nuclei;
Many-Body Physics with the Strong Interaction}, Section 2.6
(Addison-Wesley, New York, 1987).

\bibitem[6]{detar} C.E.DeTar, Phys. Rev. {D17}, 302 (1977); {\it ibid.},
p.323; Prog. Theor. Phys. {66}, 556 (1981); {\it ibid.}, p.572.


\bibitem[7]{Nambu} Y.Nambu, in {\it Preludes in Theoretical Physics},
A.deShalit, H.Feshbach, and L.vanHove, eds. (North Holland, Amsterdam,
1966).

\bibitem[8]{dRGG} A.DeRujula, H.Georgi, and S.L.Glashow, Phys. Rev.
{D12}, 147 (1975).

\bibitem[9]{Schnitzer} H.J.Schnitzer, Phys. Rev. Lett. {35}, 1540 (1975).

\bibitem[10]{Liberman} D.A.Liberman, Phys. Rev. {D16}, 1542 (1977).

\bibitem[11]{Russ}
V.G.Neudatchin, Y.F.Smirnov, and R.Tamagaki, Prog.
Theor. Phys. {58}, 1072 (1977); I.T.Obukhovsky, V.G.Neudatchin,
Y.F.Smirnov, and Y.M.Tchuvil'sky, Phys. Lett. {88B}, 231 (1979).

\bibitem[12]{Harvey} M.Harvey, Nucl. Phys. {A352}, 326 (1981).

\bibitem[13]{FFLS} A.Faessler, F.Fernandez, G.L\"{u}beck, and K.Shimizu,
Nucl. Phys. {A402}, 555 (1983).

\bibitem[14]{HLL} M.Harvey, J.Letourneaux, and B.Lorazo,
Nucl. Phys. {A424}, 428 (1984).

\bibitem[15]{WS} C.S.Warke and R.Shanker, Phys. Rev. {C21}, 2643 (1980).

\bibitem[16]{OY} M.Oka and K.Yazaki, Phys. Lett. 90B, 41 (1980).

\bibitem[17]{rib} J.E.T.Ribeiro, Z. Phys. {C5}, 27 (1980).

\bibitem[18]{Koike} Y.Koike, Nucl. Phys. {A454}, 509 (1986).

\bibitem[19]{flop} K.Yazaki, Nucl. Phys. {A416}, 87c (1984);
M.Oka and C.J.Horowitz, Phys. Rev. {D31}, 2773 (1985);
F.Lenz, J.T.Londergan, E.J.Moniz, R.Rosenfelder, M.Stingl, and K.Yazaki,
Ann. Phys. (NY) {170}, 65 (1986); Yu.A.Kuperin, S.B.Levin and Yu.B.Melnikov,
``Generalized String-Flip Model for Quantum Cluster Scattering", Department of
Mathematics and Computational Physics, St.Petersburg University report, Sec.5
(1992).

\bibitem[20]{CK} Z.-J.Cao and L.S.Kisslinger, Phys. Rev. C40, 1722 (1989).

\bibitem[21]{Shim} K.Shimizu, Rep. Prog. Phys. {52}, 1 (1989).

\bibitem[22]{MI} K.Maltman and N.Isgur, Phys. Rev. D29, 952 (1984).

\bibitem[23]{glennys} See for example J.F.Gunion, S.J.Brodsky and
R.Blankenbecler, Phys. Rev. D8, 287 (1973); S.J.Brodsky and G.R.Farrar, Phys.
Rev. D11, 1309 (1975); G.P.Lepage and S.J.Brodsky, Phys. Rev. {D22}, 2157
(1980); G.Farrar, Phys. Rev. Lett. 53, 28 (1984); S.J.Brodsky and G.P.Lepage,
in {\it Perturbative Quantum Chromodynamics}, ed. A.Muller (World Scientific,
1989).

\bibitem[24]{Botts} J.Botts and G.Sterman, Nucl. Phys. B325, 62 (1989);
J.Botts, Nucl. Phys. B353, 20 (1991).

\bibitem[25]{ILS} N.Isgur and C.H.Llewellyn-Smith, Phys. Lett. B217, 535
(1989).

\bibitem[26]{BS} T.Barnes and E.S.Swanson, Phys. Rev. D46, 131 (1992).

\bibitem[27]{BSW} T.Barnes, E.S.Swanson and J.Weinstein, Phys. Rev. D46,
4868 (1992).

\bibitem[28]{othermes} See for example D.Blaschke and G.R\"opke, Phys. Lett.
B299, 332 (1993); J.Weinstein and N.Isgur, Phys. Rev. D41, 2236 (1990);
B.Masud, J.Paton, A.M.Green and G.Q.Liu, Nucl. Phys. A528,
477 (1991); and references cited therein.

\bibitem[29]{Swanson} E.S.Swanson,
Ann. Phys. (NY) 220, 73 (1992).

\bibitem[30]{DSB} K.Dooley, E.S.Swanson, and T.Barnes, Phys. Lett. 275B, 478
(1992); K.Dooley, University of Toronto Ph.D. thesis (1993); G.Karl, ``Exotica,
Survey of Exotic Mesons", in
Proceedings of the Second Biennial Conference on
Low Energy Antiproton Physics ``LEAP '92" (Courmayeur, Italy 14-19 September
1992).

\bibitem[31]{BSKN} T.Barnes and E.S.Swanson, MIT / ORNL report MIT-CTP-2169,
ORNL-CCIP-92-15 (December 1992), submitted to Phys. Rev. C.

\bibitem[32]{Maltman} K.Maltman, Nucl. Phys. A438, 669 (1985).

\bibitem[33]{Gross} F.Gross, personal communication.

\bibitem[34]{SH} Y.Suzuki and K.T.Hecht, Phys. Rev. C27, 299 (1983).

\bibitem[35]{VPI} R.A.Arndt, L.D.Roper, R.L.Workman and M.W.McNaughton,
Phys. Rev. D45, 3995 (1992).

\bibitem[36]{DX} F.J.Dyson and N.-H.Xuong, Phys. Rev. Lett. 13, 815 (1964).

\bibitem[37]{Seth}
M.P.Locher, M.E.Sainio
and A.Svarc, Adv. Nucl. Phys. 17, 47 (1986);
K.K.Seth, pp.37-68,
Proceedings of the Third Workshop on Perspectives
in Nuclear Physics at Intermediate Energies (Trieste, 1987), ed. S.Boffi
{\it et al.} (World Scientific, Singapore).


\bibitem[38]{PDG86} M.Aguilar-Benitez {\it et al.}, Phys. Lett. 170B, 1 (1986).

\bibitem[39]{KF} T.Kamae and T.Fujita, Phys. Rev. Lett. 38, 471 (1977).

\bibitem[40]{CGMR} M.Cveti\v{c}, B.Golli, N.Manko\v{c}-Bor\v{s}tnik
and M.Rosina,
Phys. Lett. 93B, 489 (1980).

\bibitem[41]{Ryan} B.A.Ryan {\it et al.}, Phys. Rev. D3, 1 (1971).

\bibitem[42]{Ank} C.M.Ankenbrandt {\it et al.}, Phys. Rev. 170, 1223 (1968).

\bibitem[43]{Clyde} A.R.Clyde {\it et al.}, report UCRL-16275, unpublished
(1966).

\bibitem[44]{Allaby} J.V.Allaby {\it et al.}, Nucl. Phys. B52, 316 (1973).

\bibitem[45]{Nagy} E.Nagy {\it et al.}, Nucl. Phys. B150, 221 (1979).

\bibitem[46]{Breakstone} A.Breakstone {\it et al.},
Phys. Rev. Lett. 54, 2180 (1985).

\bibitem[47]{CKO} L.A.Copley, G.Karl and E.Obryk, Nucl. Phys. B13, 303 (1969).

\bibitem[48]{IK} N.Isgur and G.Karl, Phys. Rev. D20, 1191 (1979).

\bibitem[49]{KI} R.Koniuk and N.Isgur, Phys. Rev. D21, 1868 (1980).

\bibitem[50]{HI} C.Hayne and N.Isgur, Phys. Rev. D25, 1944 (1982).

\bibitem[51]{LC} Z.Li and F.E.Close, Phys. Rev. D42, 2207 (1990).

\bibitem[52]{GI} S.Godfrey and N.Isgur, Phys. Rev. D32, 189 (1985).

\bibitem[53]{CI} S.Capstick and N.Isgur, Phys. Rev. D34, 2809 (1986).

\bibitem[54]{UA4} D.Bernard {\it et al.} (UA4 collaboration), Phys. Lett. B198,
583 (1987); P.V.Landshoff, CERN report CERN-TH.6277/91.

\bibitem[55]{Roger} R.J.N.Phillips (personal communication).

\bibitem[56]{Roberts} R.G.Roberts (personal communication).

\end{references}
\end{document}